# Discrete Fourier Transform Method for Discrimination of Digital Scintillation Pulses in Mixed Neutron-Gamma Fields

M.J. Safari, F. Abbasi Davani, H. Afarideh, S. Jamili, E. Bayat

*Abstract—* **A Discrete Fourier Transform Method (DFTM) for discrimination between the signal of neutrons and gamma rays in organic scintillation detectors is presented. The method is based on the transformation of signals into the frequency domain using the sine and cosine Fourier transforms in combination with the discrete Fourier transform. The method is largely benefited from considerable differences that usually is available between the zero-frequency components of sine and cosine and the norm of the amplitude of the DFT for neutrons and gamma-ray signals. Moreover, working in frequency domain naturally results in considerable suppression of the unwanted effects of various noise sources that is expected to be effective in time domain methods. The proposed method could also be assumed as a generalized nonlinear weighting method that could result in a new class of pulse shape discrimination methods, beyond definition of the DFT. A comparison to the traditional Charge Integration Method (CIM), as well as the Frequency Gradient Analysis Method (FGAM) and the Wavelet Packet Transform Method (WPTM) has been presented to demonstrate the applicability and efficiency of the method for real-world applications. The method, in general, shows better discrimination Figure of Merits (FoMs) at both the low-light outputs and in average over the studied energy domain. A noise analysis has been performed for all of the abovementioned methods. It reveals that the frequency domain methods (FGAM and DFTM) are less sensitive to the noise effects.**

## I. Introduction

SCINTILLATORS, especially in their liquid organic form, have proved to be very useful in the detection and spectroscopy of fast neutrons. This superiority against other detectors is mainly due to their capability to discriminate between pulse shapes resulting from different particle types. As almost every neutron field has a considerable amount of gamma rays, we usually need a great deal of work to accurately account for the gamma-ray contribution to the recorded data. The contribution of gamma rays in the neutron field usually would result in complication of neutron spectroscopy, especially at lower energies. The problem is a consequence of the fact that the low-light signals could hardly be quantified, result in poor evaluations and incorrect categorization.

Digital Signal Processing (DSP), which have been exploited in recent years, poses a great possibility to have more sophisticated Pulse Shape Analysis (PSA) methods which are supposed to result in more powerful discrimination qualities. It is expected that more robust pulse shape discrimination (PSD) methods would emerge from in-depth analysis of the digitized signals, far from the possibilities of analogue techniques.

Soon after the introduction of the DSP it became obvious that transforming into the frequency domain can result in more stable behavior of signals which certainly results in better signal to noise ratios. Development of fast and powerful Discrete Fourier Transformation (DFT) algorithms have accelerated and widened such usages. Although the DFT has been previously used [1] in a simple form for analysis of liquid scintillation detectors, there is still more to be explored. The previously reported Frequency Gradient Analysis Method (FGAM) was mostly based on two factors: the smoothness of the frequency domain data and then the combination of this quality to the Pulse Gradient Analysis Method (PGAM) [2], that is indeed a time-domain method.

Here, a new discrimination method is proposed based on a proper combination of the zero-frequency component of the Discrete Sine and Cosine Transforms (DST and DCT); and the norm of a slightly modified version of the DFT. An experimental setup was arranged to get the digitized signals from a PMT coupled to a BC501 liquid scintillator. The method was thoroughly verified via analysis of the recorded signals and comparison against three other studied methods, namely the Charge Integration Method (CIM) [3], FGAM [1, 4] and the Wavelet Packet Transform Method (WPTM) [5-7]. The method shows promising properties in discrimination of neutrons and gamma rays with good Figure of Merit (FoM) values.

The manuscript is arranged as follows: Section II provides details of the proposed discrimination feature; Section III describes the experimental setup that has been arranged to obtain the digitized signals of a liquid scintillator; Section IV is mainly devoted to qualitative presentation of the results and their comparison, along with the detector calibration

M.J. Safari, H. Afarideh, E. Bayat and S. Jamili are with the Department of Energy Engineering and Physics, Amir Kabir University of Technology, Tehran, Iran. (email: mjsafari@aut.ac.ir; hafarideh@aut.ac.ir; ebayat@aeoi.org.ir; saeedjamily@gmail.com).

F. Abbasi Davani is with the Radiation Application Department, Shahid Beheshti University, Tehran, Iran. (fabbasi@sbu.ac.ir)

procedure. A detailed discussion about the analysis of FoMs and making a comparison between various methods will be presented in Section V. Section VI presents a noise analysis to compare various PSD methods. Finally, Section VII will address some concluding remarks about the manuscript's obtained results.

## II. PULSE PROCESSING METHOD

A digitized signal represented by a vector of *N*-element data, $x_k$, (Fig. 1) contains essential information about the type of particle. However, it is anticipated that transforming into the frequency domain would result in a better understanding of the constituent (frequency) components of the signal which might be indicative for classification purposes. This conclusion is rooted in the fact that transforming into frequency domain usually would reduce the effects of noise and other unwanted perturbative/uncorrelated effects [1, 4, 8]. This lower sensitivity is mainly due to the fact that the frequency domain data are based on collecting various frequency contributions that would act as a noise reducer, much like the charge integration but here in the frequency domain. For example, Liu et al [1], and later Jun et al [4] have presented the FGAM, which almost entirely was based on conversion of the older PGAM [2] into the frequency space. However, other authors, preferred not to use the Fourier transform, and instead used other frequency domain methods, like the Wavelet Transform Method (WTM) by Yousefi et al. [8] or the Power Spectrum Analysis Method (PSAM) by Luo et al. [9]. These studies seem far from complete and as such, need more room for further exploration of the DFTM capabilities.

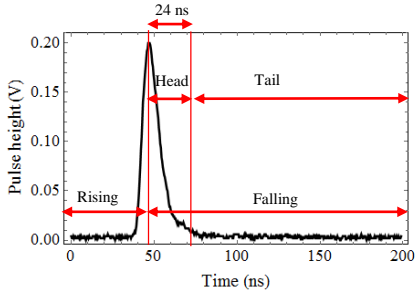

Fig. 1. A typical digitized pulse in the time domain.

### A. Gatti's Linear Weighting Method

A more general basis for PSD has been introduced by Gatti and De Martini [10] which demands finding an optimized observable (*i.e.*, the PSD feature) by means of a linear weighting

$$\text{PSD} = \sum_k w_k x_k, \quad (1)$$

in which the specific form of the optimized weight ($w_k$) should be determined appropriately. Due to its generality, most linear methods could be considered as a special case of this formula. For example, the CIM which integrates on the tail region of the falling portion of the signal (see section IV.C.1 for more details) is based on a step weight function [11]

$$w_k = \begin{cases} 0, & k \notin \text{Tail}, \\ 1, & k \in \text{Tail}. \end{cases}$$

### B. DFTM

The well-known constituents of the DFT are the DST and DCT [12]. This transformation is interesting in the DSP community and has useful properties for analysis of the scintillation signals. Here, the interesting feature is that there is a lag between sine and cosine transforms (this is shown in Fig. 2a). One could represent the more conventional Fourier transform as

$$\hat{x}_k = \frac{1}{N^{(1-a)/2}} \sum_{n=1}^{N} x_n \exp\left[\frac{2\pi i b}{N}(n-1)(k-1)\right]. \quad (2)$$

A special case of this transformation could be deduced by setting $a=1$ and $b=-1$ in Eq. (2) which is found to be suitable for our signal processing studies. The norm of this slightly modified DFT would be combined with DST and DCT to define a new PSD feature as it follows. Taking the exponential function (restricted to the non-negative values of $t$) as

$$f(x) = \begin{cases} e^{-\lambda t}, & t \geq 0, \\ 0, & t < 0, \end{cases}$$

the norm of the Fourier transform would be as follows

$$\hat{f}(\omega) = \frac{1}{\sqrt{2\pi}} \frac{1}{\omega + \lambda},$$

which is a monotonically decreasing function.

Scintillation signals are of exponential nature which could be approximated by several heuristic model functions [13]. DST/DCT of such pulses would result in an oscillating trend in the frequency domain. Fig. 2a represents the DST and DCT of the typical signal of Fig. 1, and Fig. 2b shows the real and imaginary parts of the DFT which their shape remains almost unchanged for low- and high-energy signals. By taking the norm of this latter transformation one gets a smooth pattern in the frequency space (Fig. 2c).

The proposed discrimination feature is based on evaluation and combination of the total of the norm of the DFT and zero-frequency component of DST and DCT according to the following relation

$$\text{PSD} = \frac{\sum_k \|\hat{x}_{\text{DFT},k}\|^2}{\hat{x}_{\text{DST},1} \cdot \hat{x}_{\text{DCT},1}}, \quad (3)$$

for the signal $\hat{x}$. Noting the Parseval identity and after a little algebra we find the following equivalent form

$$\text{PSD} = \frac{\sum_{j,k} x_j x_k \delta_{ik}}{\sum_{j,k} x_j x_k \cos\left(\frac{\pi j}{2N}\right) \sin\left(\frac{\pi k}{2N}\right)}. \quad (4)$$

In Gatti's sense [10, 11], it could be addressed as

$$\text{PSD} = \frac{\sum_{j,k} x_j x_k w_{jk}}{\sum_{j,k} x_j x_k u_{jk}},$$

where

$$w_{jk} = \delta_{jk},$$
$$u_{jk} = \cos\left(\frac{\pi j}{2N}\right)\sin\left(\frac{\pi k}{2N}\right).$$

This is a nonlinear generalization of Gatti's method which benefits from the correlation of the signal. It is demanded that this property would reduce sensitivity to the random noise.

We empirically found that dividing this quantity by the total charge of the pulse, enhances discrimination factor, so Eq. (3) was modified as

$$\text{PSD} = \frac{\sum_{k \in \text{Falling}} \left\|\hat{x}_{\text{DFT},k}\right\|^2}{\hat{x}_{\text{DST},1} \cdot \hat{x}_{\text{DCT},1}} \frac{1}{Q_{\text{Tot}}},$$

where we have

$$Q_{\text{Tot}} = \sum_k x_k.$$

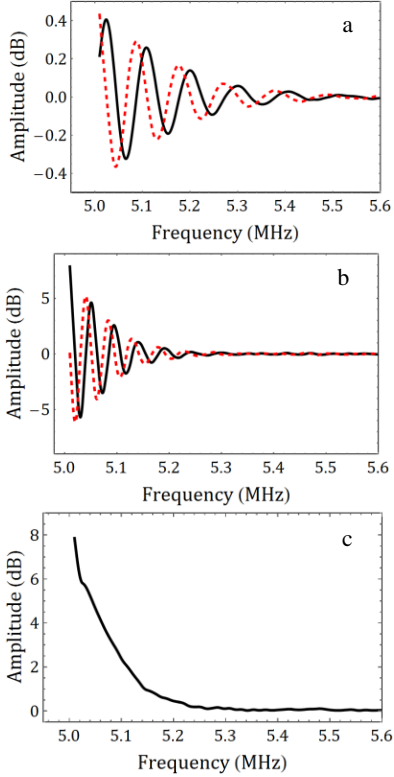

Fig. 2. (a) Fourier sine (full line) and cosine (dotted line) transforms for a typical signal. (b) Real (full line) and imaginary parts (dotted line) of the DFT for the typical signal. (c) Norm of the amplitude of imaginary and real parts of DFT signal.

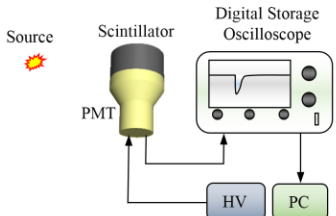

Fig. 3. A schematic representation the experimental setup.

Usually the PSD methods are sensitive to selected portion of the signal being analyzed, *viz.* the full-length signals or the falling portion of the signal (starting at the local maximum value). In our studies, it is found that taking the falling portion is suitable for neutron-gamma discrimination purposes.

### III. EXPERIMENTAL SET-UP

A cylindrical ⌀3"×3" BC501 scintillator encapsulated in an aluminum cell was used as the detector, which was coupled to a Hamamatsu R6091 3" photomultiplier tube (PMT). The cell boundaries were painted with $TiO_2$ white reflector to allow for better light collection efficiency. It also should be added that the PMT has been thoroughly wrapped in a *μ*-metal sheath to suppress disturbances resulting from external electromagnetic effects. The PMT was operated at 1700 V, and the output of its anode was directly fed into a digital storage oscilloscope which is capable of digitizing and storing the incoming signals at a sample rate of 5 GHz with 8-bit precision. The detector was irradiated by a 100 mCi Am-Be neutron/gamma source. The source was located at 50 cm distance from the curved boundary of the detector, to yield a slightly larger count rate. The simulation study suggests that the source would create a flux of about 180 #/$cm^2$-s at the detector location. A schematic presentation of experimental configuration is shown in Fig. 3.

### IV. RESULTS AND DISCUSSION

#### A. Energy Calibration

An organic scintillation detector is typically calibrated using the radioactive isotopes which emit gamma rays with known energies, and localizing of the Compton edge position in the measured spectra. Practically the anticipated spectrum of an organic scintillator would not show a photopeak, mainly because of its low-Z material. The response is usually an asymmetric Compton continuum with a local maximum and a broadened trend that is dependent on the energy resolution of the detector. The most challenging problem in accurate calibration is to correctly specify the Compton edge location. The literature is quite diverse about the location of the Compton edge with respect to the local maximum of the pulse height distribution. The values range from: 66% (by Beghian et al [14]), 70% (by Honecker and Grässler [15]), 85% to 88% (by Bertin et al [16]); 89% (by Knox and Miller [17]) and 78% to 82% (by Swiderski et al [18]). This difference in Compton edge location could be attributed to the effects of multiple Compton scattering that is directly related to the detector's size. Although, Dietze and Klein [19, 20] have introduced a calibration method which reduces the uncertainty in precise determination of the edge position, which is supposed to be mainly related to apparatus- and laboratory-dependent factors. This method is based on a Monte Carlo simulation of the response function rather than *a priori* assumption of the Compton edge position. To do such a comparison, one should adjust the *x*-axis channel values to find the best conformity between data sets; noting that, both the experimental and simulation data should properly be normalized (*e.g.*, to the maximum value) prior to the analysis. One should appropriately smear out the simulated response function with the experimental energy resolution to find the best similarity between both data. The smearing could be performed by a convolving algorithm taking the resolution of the detector as its response [21].

The gamma-ray lines of several radionuclide sources ($^{22}$Na, $^{137}$Cs, $^{60}$Co, $^{152}$Eu) were used for calibration. The first two sources are detailed below. The activity of $^{22}$Na ($E_\gamma$=0.511 MeV, $E_\gamma$=1.28 MeV) was 2 $\mu$Ci, and the $^{137}$Cs ($E_\gamma$=0.662 MeV) was 0.5 $\mu$Ci. These sources were almost localized points with small plexiglass backings. The location of sources at the time of measurement were set as close as possible to the center of the lateral curved outer boundary of the detector.

The details of geometry and material of the source and detector have been implemented into the FLUKA transport simulation code [22]. After a thorough simulation study of the response of the detector, we followed Dietze and Klein [19] to introduce a linear interpolation between the light output channels and the corresponding electron equivalent energy (in keV$_{ee}$)

$$E = aC - b, \quad (5)$$

where $C$ and $E$ would be regarded as the light output channel and the corresponding electron equivalent energy, respectively, and $a$=54.79±1.21, $b$=18.08±1.10. Fig. 4 shows this comparison between the measured and the simulated data for $^{137}$Cs and $^{22}$Na sources. The proposed method by Dietze and Klein [19] for calibration of organic scintillators, not only defines the correct location of the Compton edge, but also simultaneously leads to an estimate of the resolution of the detector. This technique relies on comparing the smeared Monte Carlo response against the experimental response, to find the best possible conformity between them. It is useful to point out that the *conformity test* could be quantified by a correlation comparison.

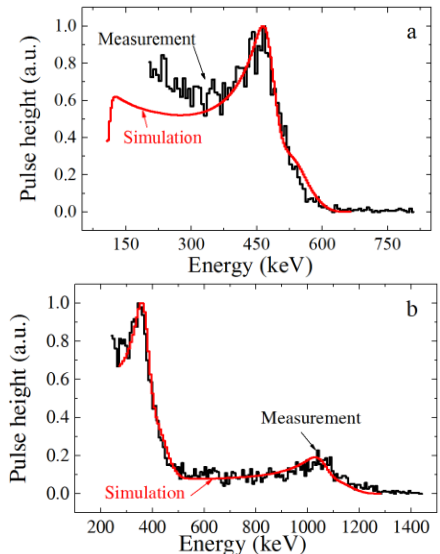

Fig. 4. Response of the detector to the gamma rays of (a) $^{137}$Cs and (b) $^{22}$Na calibration sources, from the simulation and the measurement.

### B. The PSD Results

The data from the digital oscilloscope were stored in separate files for later offline analysis. To study the proposed discrimination method, the PSD factor from Eq. (3) was applied to each one of the output signals. Plotting the PSD factor against the total light output of the signal (*i.e.*, sum of the signal) a separated pattern was found (see Fig. 5) which is attributed to neutrons and/or gamma rays. This attribution would later be confirmed in comparison to other validated methods for PSD. Although, the results are comparable from around 100 keV$_{ee}$ up to about 2000 keV$_{ee}$, we have focused on the light outputs between 100 keV$_{ee}$ – 1600 keV$_{ee}$ (equivalent to recoiled proton energies of about 0.5 MeV$_{pe}$ – 4.4 MeV$_{pe}$ [17] where the MeV$_{pe}$ denotes the recoiled proton equivalent energy in MeV) in which all methods were able to pose a reliable behavior in their FoM properties. The next section is devoted to the analysis of results in comparison to other methods applied to the same set of recorded signals.

### C. Comparison to Other Methods

There are a wide range of digital PSD methods in the literature, showing different degrees of success in the neutron-gamma discrimination capability. We selected three methods to have a sense about the relative quality of our proposed method. The implemented methods for this purpose were the following:

- the Charge Integration Method (CIM),
- the Frequency Gradient Analysis Method (FGAM); and
- the Wavelet Packet Transform Method (WPTM).

As it is beyond the scope of this manuscript, the theoretical background of these methods will not be elaborated here. We will refer the reader to the corresponding related references that have been addressed in the following relevant subsequent subsections and limit our discussion to a brief introduction.

#### 1) Charge Integration Method

The CIM could be regarded as a standard neutron-gamma PSD method. Although different degrees of successes have been obtained using this method, it is generally regarded as the reference method of neutron-gamma discrimination, specifically in the DSP jargon which was among the first digital PSD implementations [23]. The method is based on performing two integration operations over a *short* and a *long* integration limits of the falling portion of signals. A typical scintillation signal has two parts: a *rising* and a *falling* portion respectively in which this latter part is constituted by a *head* and a *tail* (see Fig. 1). Usually in CIM we are interested in the tail portion, because its integral would give the appropriate PSD feature. For more details the reader is referred to the discussions made by Bell [23] Jordanov and Knoll [24], Yong-Hao et al [25], Nakhostin [26] and Zhonghai et al [3]. However, the discrimination quality is dependent on the optimal selection of integration limit. This problem has been studied in the literature [3, 23, 27], suggesting different values ranging from 20 ns to 30 ns. In our case we found that 24 ns is a weakly optimal choice.

*2) Frequency Gradient Analysis Method*

The frequency gradient analysis method (FGAM) [1, 4] is a natural possible development of the older pulse gradient analysis method [2, 28] into the frequency domain. The method is mainly based on evaluation of the zero frequency component of signals which differs for neutrons and gamma rays. However, its usefulness is a consequence of the extra data that had been extracted from considering another frequency component of the signals. Strictly speaking, the discrimination factor is defined as the difference between zero frequency and an optimal frequency (of $k$-index) which has the largest value of difference:

$$k_f = \frac{|\hat{x}_1 - \hat{x}_k|}{k},$$

where the $k$-index should be appropriately determined. Liu et al [1] determined the $k$-index to 4.6 MHz. The optimized $k$-index might vary for different signals. Thus, we preferred to evaluate the $k_f$ for some other frequencies, and then use the best value which gives better seperation.

*3) Wavelet Packet Transform Method*

The Wavelet Packet Transform Method (WPTM) [5-7] (also called the *simplified digital charge collection method*) is rooted in the use of a discrimination parameter that is defined as follows

$$k = -\log \sum_{n \in \text{Tail}} x_n^2,$$

where by $n \in \text{Tail}$ we mean that the summation would be accomplished for the tail region of the falling portion of each signal. It is notable that the portion of the signal that should be considered as the tail region would be optimized properly, just as it was done (but not necessarily equivalent to the value that was used) for the implementation of CIM. By means of a systematic study for optimization of the integration window, it is found that here the tail region would be about 10 ns narrower than that of the CIM.

*4) Comparison*

Each one of the abovementioned methods has been implemented in a unified manner to process a total of 80000 signals that were taken during the experiment. The resultant output of the PSD feature of each one has been stored in a matrix with respect to the corresponding total sum of the full-length signal *i.e.*, the total light output. Fig. 5 shows the plot of these data with inclusion of the guide lines showing position of the gamma rays emitted from the $^{22}$Na and $^{137}$Cs calibration sources. Moreover, a heuristic discrimination line (shown by dashed red line) has been added for more clarity in qualitative comparisons. It seems that the observed discrimination in Fig. 5 is satisfactory in qualitative terms.

We will further focus on quantitative comparison of the methods in the lower end of data, around 200 keV$_{ee}$, as well as the average FoM for the whole range of reasonably available light outputs (*i.e.*, 100 keV$_{ee}$ to 1600 keV$_{ee}$). The 200 keV$_{ee}$ region of interest is highlighted by yellow semi-transparent color in Fig. 5.

Contrary to other three methods which show scattered patterns at higher energies, DFTM exhibits more confined PSD pattern and well separated plumes, especially at higher energies. It also poses a relatively linear tendency, more obvious at higher energies. This quality will be discussed in the next section in quantitative terms.

V. ANALYSIS OF FoMs

The well-established and widely accepted method for quantitative comparison of various pulse shape discrimination techniques is to evaluate the FoM and its constitutive components. The method is to fit a double-Gaussian function, in which, each one of the Gaussian functions has a corresponding centroid and an FWHM. The FoM is defined as

$$\text{FoM} = \frac{S}{w_1 + w_2}$$

where $S$ refers to the separation between the centroids, and $w_1 + w_2$ is the sum of FWHMs. Noting the nonlinear nature of the PSD methods, one should take a differential slice around each light output. Here, the width of the differential slice has been set to 30 keV$_{ee}$.

Evaluating the FoM for each method in the 200 keV$_{ee}$ band, one could find a comparison of the discrimination quality that is supposed to be representative of the lower end of our data. Fig. 6 shows the distribution of PSD features of various methods for the region around 200 keV$_{ee}$. Overlaid in the same graphs are the double-Gaussian fits (in thick red lines) which were found by means of a linear least squares fitting method. The adequacy of a model (here the normal distribution) to fit a set of data could be verified by the several goodness-of-fit tests [29], such as Pearson $\chi^2$ and Kolmogorov-Smirnov [30]. In our case, the Pearson $\chi^2$ estimation of goodness factors were about 0.88 and the result of Kolmogorov-Smirnov tests were at least 0.82. This tests concludes that the data fits reliably enough to the normal distribution. Table 1 shows the results of these tests for left- and right-hand-side peaks of Fig. 6. The fitting procedure could also return uncertainty of the fitting parameters, to be able to evaluate the uncertainty in $S$ and $w_1 + w_2$. For estimation of the FoM uncertainty, one should combine the uncertainty of $S$ and $w_1 + w_2$ according to the rules of error propagation [31]. The same table shows $S$, $w_1$ and $w_2$ along with their corresponding uncertainties for 200 keV$_{ee}$.

In addition to the important low-energy region, one should also be worried about the FoM at higher energies. We have evaluated the FoM at different energies in the range of about 100 keV$_{ee}$ – 1600 keV$_{ee}$. Fig. 7 demonstrates the FoM for different methods at different energies. The data points were joined with straight lines to guide the eye for better comparison. It seems that the DFTM shows relatively better FoM across the energy range. According to Fig. 5, the improvement in FoM at higher energies could be attributed to the reduction of FWHMs. However, at lower energies, the well separation of plumes is responsible for improvement of FoMs.

Table 1. Main properties of the Gaussian fits, presented in Fig. 6, along with their corresponding Pearson $\chi^2$ and Kolmogorov-Smirnov goodness of fit tests.

| Method | S | $w_1$ | $w_2$ | Pearson $\chi^2$ | Kolmogorov-Smirnov |
|---|---|---|---|---|---|
| CIM | 3.52E-1±1.8E-3 | 1.08E-1±2.5E-4 | 1.91E-1±4.5E-4 | 0.91 | 0.94 |
| FGAM | 1.34E-2±1.2E-6 | 4.47E-3±1.1E-5 | 7.30E-3±1.7E-5 | 0.90 | 0.88 |
| WPTM | 1.10E+0±8.2E-3 | 5.18E-1±1.2E-4 | 2.94E-1±6.9E-4 | 0.96 | 0.87 |
| DFTM | 4.19E+0±2.6E-2 | 1.05E+0±2.5E-3 | 1.77E+0±4.2E-3 | 0.96 | 0.86 |

To summarize the results, a comparison of FoMs at the low light output region (*i.e.*, 200 keV$_{ee}$), as well as the averaged FoM over the whole range of study (100 keV$_{ee}$ – 1600 keV$_{ee}$) is reported in Table 2. This comparison shows that the DFTM results in better discrimination at 200 keV$_{ee}$, as well averaged over the energy range of study.

Due to the limited bit-precision and consequently our limited dynamic range, we concentrated on the mid-energy part of the Am-Be spectrum (below 1600 keV$_{ee}$). However, from Fig. 7 one could conclude that there would be no significant increase in FoM beyond 1600 keV$_{ee}$. Although, it needs careful experimental analysis using ADCs with at least 10-bit resolution to precisely determine the variation of FoMs at those energies [11].

Table 2. Comparison between various methods regarding their FoMs and CPU usage.

| Method | FoM at 200 keV$_{ee}$ | Average FoM | CPU usage ($\mu$s/pulse) |
|---|---|---|---|
| CIM | 1.18±0.02 | 1.55±0.02 | 244 |
| FGAM | 1.14±0.01 | 1.49±0.03 | 123 |
| WPTM | 1.35±0.01 | 1.52±0.02 | 229 |
| DFTM | 1.49±0.01 | 1.66±0.02 | 243 |

## VI. NOISE ANALYSIS

Extra information could be obtained about affections of the random noise (*i.e.*, the quality of signals) on the FoM. In brief, it could be said that the more robust against the noise, the more suitable for high-noise conditions. We added a fictitious random noise to each signal, meaning that the recorded voltage by the ADC is indeed a mean value. Fig. 8 shows a signal with a 2 mV extra random noise (red line), keeping the original signal as its average (black line). It is obvious that a low-energy signal would show more reaction to the added noise. We select the 200 keV$_{ee}$ region for quantitative comparison of FoMs. The results are presented in Fig. 9 and shows that the frequency domain methods (*i.e.*, DFTM and FGAM) are less sensitive to the noise effects. Contrary to the common sense, it reveals that the FoM is not monotonically decreasing with respect to the amplitude of the noise. However, one can find that the change in shape of frequency domain methods is reasonably similar. Equivalently, this is correct also for the time-domain methods, meaning that CIM and WPTM are of almost similar behavior.

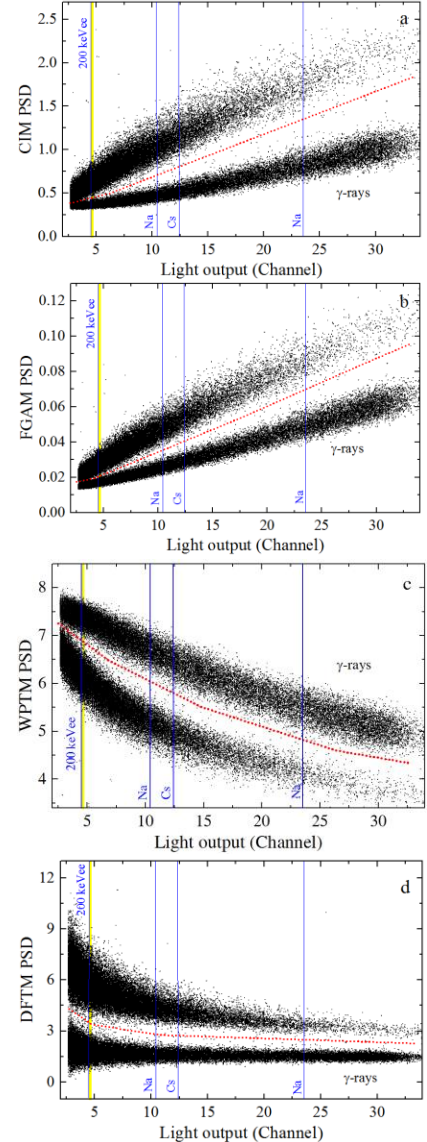

Fig. 5. PSD performance of various methods: (a) CIM, (b) FFTM, (c) WPTM, (d) DFTM. The horizontal axis is the total light output of the signal in units of channels, related to the energy by Eq. (5).

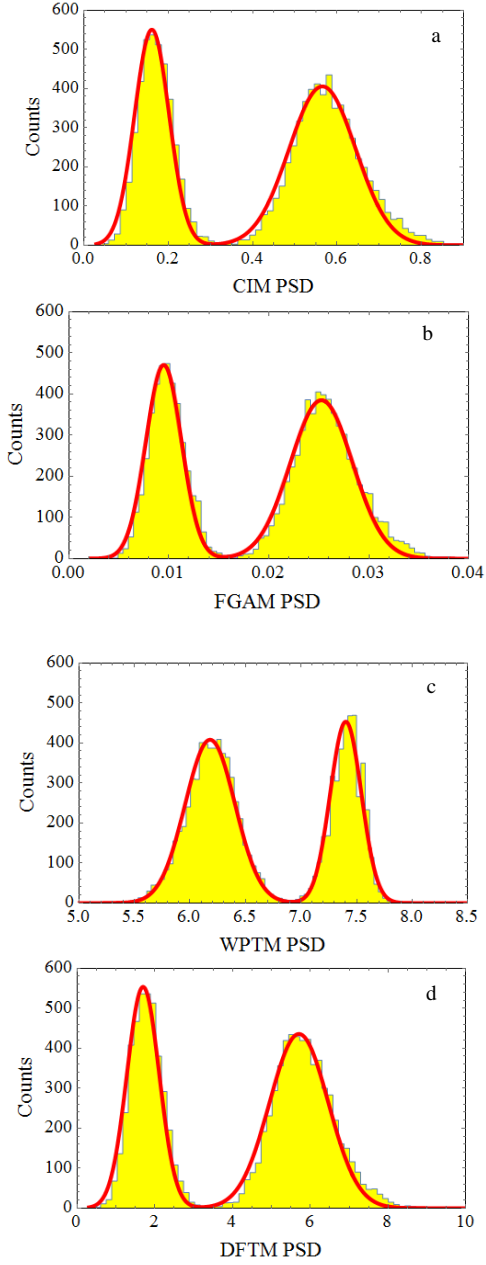

Fig. 6. Evaluation of FoM for 200 keV$_{ee}$ (or 1.2 MeV$_{pe}$) pulses as resulted from various discrimination methods: (a) CIM, (b) FGAM, (c) WPTM, (d) DFTM. The histogram (in yellow) shows the original data, and the smooth thick line (in red) shows the double-Gaussian fit to the data.

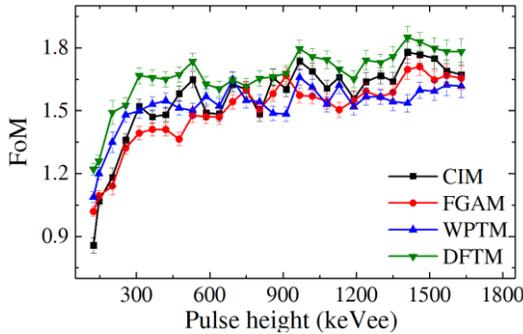

Fig. 7. Comparison of FoM at different light outputs (pulse heights).

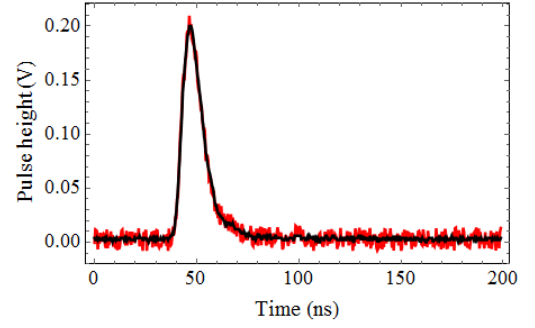

Fig. 8. A signal with 2 mV fictitious added random noise.

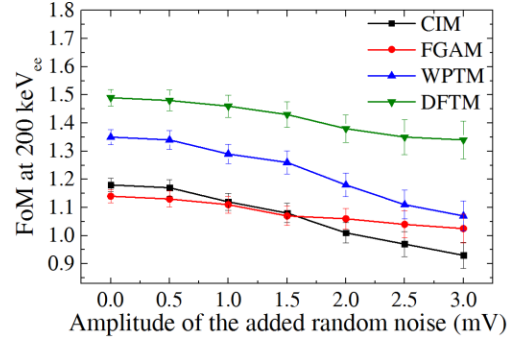

Fig. 9. Variation of FoM at 200 keV$_{ee}$ with respect to the amplitude of the added random noise.

## VII. CONCLUSIONS

A new method for discrimination of the digitized neutron-gamma signals has been developed based on a combination of the discrete Fourier transform (DFT), the discrete sine transform (DST) and the discrete cosine transform (DCT). It brings a new look at the frequency domain techniques for more sophisticated analysis of signals in the scintillation spectroscopy. The method, along with some other methods, namely the CIM, FGAM and WPTM have been applied to the digitized signals of an experimental setup. The experimental arrangement was based on feeding the output current from PMT anode coupled to a BC501 scintillator into a digital storage oscilloscope which could store signals at the sample rate of 5 GHz with 8-bit precision. A total of 80000 signals resulting from an Am-Be neutron/gamma source were studied. It has been observed that the method is capable of discriminating the neutron and gamma-ray signals of an Am-Be source for the energy range of 100 keV$_{ee}$ to 1600 keV$_{ee}$. Undoubtedly, this dynamic range is limited by the 8-bit precision of our digitizer. Any change in the setup to gather higher-quality signals, or increasing precision of the digitizer would directly affect the discrimination FoMs. There are reports showing that 10-bit is an optimal choice for resolution of the ADC [11]. Moreover, other factors such as the stray capacitance (mainly due to the coaxial cable), quality of the scintillator itself, reflector painting, detector assembly, and voltage divider could affect the quality of signals, directly influencing the obtained PSD pattern.

The DFTM shows average FoM of about 1.66±0.02 in comparison to the other studied methods (1.55±0.02, 1.49±0.03 and 1.52±0.02 for the CIM, FGAM and WPTM, respectively). For the current set of signals, the DFTM shows better performance.

A noise analysis could be performed by adding a fictitious random noise with a specific amplitude to the signals. This would simulate a reduction in the resolution of the ADC. Such a study was conducted for 200 keV$_{ee}$, confirming that the frequency domain methods are less sensitive to the noise.

The DFTM needs evaluation of the DFT, DST and DCT of the recorded signals which could be calculated by state-of-the-art programming languages (such as Fortran and C++, especially using the third-party libraries like IMSL), and numerical/mathematical softwares (such as Mathematica and MATLAB). The CPU usage of different methods is highly dependent on the method implementation. The last column of Table 2 shows the wall-clock computation time on a 2.8 GHz Intel Core i7 PC for our set of signals. While in our implementation, the fastest method is FGAM with computation time of about 123 $\mu$s per pulse, other methods need at least 230 $\mu$s per signal. The comparability of running-time in our offline implementations argues that the DFTM would also fit for field-work real-time applications. Nonetheless, such applications demanding appropriately programmed systems (such as FPGA) to perform the required numerical operations.

ACKNOWLEDGMENTS

This first author wishes to thank Dr. S. M. Ayyoubzadeh for his valuable discussions and comments.